\documentclass[twocolumn]{aastex62}
\usepackage{natbib}
\usepackage{amsmath}

\newcommand{\teff}{$T_{\text{eff}}$}
\newcommand{\logg}{$\log g$}
\newcommand{\msun}{$M_{\odot}$}

\shorttitle{WD COMPANIONS OF BSS}
\shortauthors{GOSNELL ET AL.}

\begin{document}

\title{Constraining Mass Transfer Histories of Blue Straggler Stars with COS Spectroscopy of White Dwarf Companions}

\author[0000-0002-8443-0723]{Natalie M. Gosnell}
\affiliation{Department of Physics, Colorado College, 14 E. Cache La Poudre St, Colorado Springs, CO 80903, USA}
\email{ngosnell@coloradocollege.edu}

\author[0000-0002-3944-8406]{Emily M. Leiner}
\affiliation{Department of Astronomy, University of Wisconsin - Madison, 475 N. Charter Street, Madison, WI 53706, USA}
\affiliation{Center for Interdisciplinary Exploration and Research in Astrophysics (CIERA) and Department of Physics and Astronomy, Northwestern University, 2145 Sheridan Rd, Evanston, IL  60208, USA}

\author{Robert D. Mathieu}
\affiliation{Department of Astronomy, University of Wisconsin - Madison, 475 N. Charter Street, Madison, WI 53706, USA}

\author[0000-0002-3881-9332]{Aaron M. Geller}
\affiliation{Center for Interdisciplinary Exploration and Research in Astrophysics (CIERA) and Department of Physics and Astronomy, Northwestern University, 2145 Sheridan Rd, Evanston, IL  60208, USA}
\affiliation{Adler Planetarium, Department of Astronomy, 1300 S. Lake Shore Drive, Chicago, IL 60605, USA}

\author[0000-0002-1116-2553]{Christian Knigge}
\affiliation{School of Physics and Astronomy, University of Southampton, Highfield, Southampton, SO17 IBJ, UK}

\author[0000-0003-3551-5090]{Alison Sills}
\affiliation{Department of Physics and Astronomy, McMaster University, 1280 Main St. W, Hamilton, ON  L8S 4M1, Canada}

\author{Nathan W. C. Leigh}
\affiliation{Departamento de Astronom\'ia, Facultad de Ciencias F\'isicas y Matem\'aticas,
Universidad de Concepci\'on, Concepci\'on, Chile}
\affiliation{Department of Astrophysics, American Museum of Natural History, Central Park West and 79th Street, New York, NY 10024, USA}


\begin{abstract}
Recent studies show that the majority of blue straggler stars (BSSs) in old open clusters are formed through mass transfer from an evolved star onto a main-sequence companion, resulting in a BSS and white dwarf (WD) in a binary system. We present constraints on the mass transfer histories of two BSS-WD binaries in the open cluster NGC 188, using WD temperatures and surface gravities measured with \textit{HST} COS far-ultraviolet spectroscopy. Adopting a \textit{Gaia}-based cluster distance of $1847\pm107$ pc, we determine that one system, WOCS 4540, formed through Case C mass transfer resulting in a CO-core white dwarf with \teff=$17000^{+140}_{-200}$ K and a \logg=$7.80^{+0.06}_{-0.06}$, corresponding to a mass of $0.53^{+0.03}_{-0.03}$ \msun\ and a cooling age of $105^{+6}_{-5}$ Myr. The other system, WOCS 5379, formed through Case B mass transfer resulting in a He-core white dwarf with \teff=$15500^{+170}_{-150}$ K and a \logg=$7.50^{+0.06}_{-0.05}$, corresponding to a mass of $0.42^{+0.02}_{-0.02}$ \msun\ and an age of $250^{+20}_{-20}$ Myr. The WD parameters are consistent across four different cluster distance assumptions. We determine possible progenitor binary systems with a grid of accretion models using MESA, and investigate whether these systems would lead to stable or unstable mass transfer. WOCS 4540 likely resulted from stable mass transfer during periastron passage in an eccentric binary system, while WOCS 5379 challenges our current understanding of the expected outcomes for mass transfer from red giant branch stars. Both systems are examples of the value in using detailed analyses to fine-tune our physical understanding of binary evolutionary processes.
\end{abstract}

\keywords{binaries: close, blue stragglers, open clusters and associations: individual (NGC 188), stars: evolution, white dwarfs}

\section{INTRODUCTION}

Thorough studies of old open clusters reveal a variety of stellar populations that do not follow single-star evolutionary pathways \citep[e.g.,][]{Landsman1997,Geller2017,Mathieu2019}. 
Stars that fall in unexpected areas of an optical color-magnitude diagram (CMD) or a Hertzsprung-Russell (HR) Diagram have histories that have altered their stellar temperatures, luminosities, or both. In evolved open clusters, these alternative pathway stellar products can make up to 25\% of the total evolved stellar population \citep{Mathieu2015,Gosnell2015}. In open clusters, many of these stars are categorized as blue straggler stars (BSSs).

Classical BSSs are traditionally empirically defined to be stars bluer and brighter than the main-sequence turnoff \citep{Sandage1953}, although BSS analogs called ``blue lurkers" can be found on the main sequence \citep{Leiner2019}. BSSs are found in many stellar populations, including in globular clusters \citep[e.g.,][]{Piotto2004,Knigge2008} and in the field as blue metal-poor stars \citep{Preston2000,Carney2001}. In order to appear in this region of a CMD, enough mass must be added to a main-sequence star such that it exceeds the current main-sequence turnoff mass. In old open clusters, this process happens primarily 
through binary mass transfer \citep{Geller2011,Gosnell2015}. It is also possible to create a binary BSS through mergers in triple systems, such as those driven by the Lidov-Kozai mechanism \citep{Perets2009,Naoz2014}, and through binary-mediated collisions in dynamically-active environments such as globular cluster cores \citep[e.g.,][]{Leigh2011}.

Across multiple open clusters and in the field, most BSSs are found in wide binaries with periods of order 1000 days. Some of these have circular orbits, in contrast to main-sequence binaries at similar periods that are typically eccentric
\citep{Carney2001,Sneden2003,Latham2007,Geller2009}. These orbital periods are consistent with the final orbital periods expected after mass transfer from an asymptotic giant branch (AGB) star onto a main-sequence companion, known as Case C mass transfer \citep{Chen2008, Gosnell2014}. This process results in a BSS with the remnant core of the giant star donor as a companion, observed as a CO-core white dwarf \citep[WD,][]{Paczynski1971}. BSSs can also form as a result of mass transfer from a red giant branch (RGB) star, known as Case B mass transfer, which results in a BSS with a He-core WD companion at a shorter binary period of order 100 days \citep{Chen2008}. 

The \textit{Gaia} mission \citep{gaiamission} continues to improve our understanding of cluster populations \citep[e.g.,][]{Choi2018} and discover new open clusters that were previously unknown \citep[e.g.,][]{Cantat-Gaudin2018}. As a result, many more stars on alternative stellar evolution pathways will be uncovered. Placing these products in context within stellar evolutionary processes requires knowing the formation mechanisms responsible for and the future evolution of these systems. In this paper we focus on the BSS population of NGC 188 in order to constrain the mass-transfer histories that are responsible for creating the majority of old open cluster BSSs. 

\subsection{The NGC 188 Blue Straggler Population}

\begin{figure}[h]
\begin{center}
\includegraphics[scale=0.35]{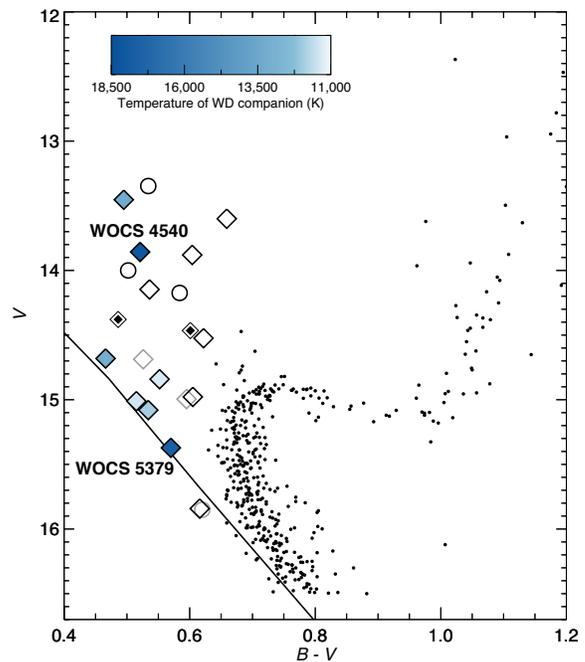}
\end{center}
\caption{From \citet{Gosnell2015}, a CMD of NGC 188 cluster members. The solid black line shows the zero-age main sequence. The BSS symbols indicate binarity (diamonds: binary, double diamond: double-lined binary, circle: non-velocity variable). BSS with photometric WD detections are filled in with a color from light to dark blue, indicating the approximate temperature of the WD companion. Hotter WD companions are younger. The two BSS in this study, WOCS 4540 and WOCS 5379, are labeled.}
\label{fig:ngc188cmd}
\end{figure}

The old open cluster NGC 188 has one of the most thoroughly studied BSS populations to date. The cluster contains 21 kinematically (radial velocity and proper motion) confirmed BSSs, the majority of which are in long-period binary systems with known periods and eccentricities \citep[and references therein]{Mathieu2009}. 
An additional UV-bright BSS in NGC 188 was detected in \citet{Subramaniam2016}, but it does not have confirmed kinematic membership. \citet{Gosnell2014} and \citet{Gosnell2015} detected WD companions of 7 kinematic member BSSs using far-ultraviolet photometry, as shown in Figure~\ref{fig:ngc188cmd}. The presence of a WD in binaries with periods appropriate for Case B or Case C mass transfer indicates that these BSS-WD systems are post-mass-transfer binaries. Taking into account WD detection limits, \citet{Gosnell2015} estimate that 67\% of the NGC 188 BSSs form through mass transfer.

The presence of a moderate-temperature WD companion in a post-mass-transfer system sets the timeline of the mass-transfer history. The WD cooling age is a measure of the time since mass transfer ended \citep{Gosnell2014}. This timeline, combined with the age and current main-sequence turnoff mass of the cluster, provides important constraints on the progenitor (pre-mass transfer) system. Possible histories for the three most recently formed BSSs in NGC 188 were presented in \citet{Gosnell2014}, but the specifics of the mass-transfer physics depend heavily on the donor-star star core mass, which becomes the WD mass. In order to better constrain the pre- and post-mass-transfer systems and to constrain the specific mass-transfer histories, in this paper we present \textit{Hubble Space Telescope} Cosmic Origins Spectrograph (COS) spectra of WD companions of two BSS systems in NGC 188: WOCS 4540 and WOCS 5379. The binary properties of both systems are provided in Table~\ref{tab:binparams}. In Section~\ref{sec:observations} we outline the observations and in Section~\ref{sec:analysis} we detail the spectral analysis that yields WD masses and cooling ages. These masses and ages inform the possible mass transfer histories explored in Section~\ref{sec:discussion}, and we provide our Summary in Section~\ref{sec:summary}.

\begin{table*}
\begin{center}
\caption{Previously known properties of WOCS 4540 and WOCS 5379} \label{tab:binparams}
\begin{tabular*}{5in}{@{\extracolsep{\fill}} l l l}
Property & WOCS 4540 & WOCS 5379 \\
\hline
Position (J2000) & 00:45:18.27, +85:19:19.85 & 00:50:10.79, +85:14:38.08 \\
$V^{\mathrm{a}}$ & 13.857 & 15.372   \\
$B-V^{\mathrm{a}}$ & 0.521 & 0.570  \\
BSS \teff$^{\mathrm{b}}$ & $ 6590\pm100$ K& $6400\pm120$ K  \\
Binary period$^{\mathrm{c}}$ &  $3030\pm70$ days & $120.21\pm0.04$ days  \\
Orbital eccentricity$^{\mathrm{c}}$ &  $0.36\pm0.07$  & $0.24\pm0.03$ \\
\hline
\multicolumn{3}{l}{$^{\mathrm{a}}$ from \citet{Sarajedini1999}   }\\
\multicolumn{3}{l}{$^{\mathrm{b}}$ from \citet{Gosnell2015}  }\\
\multicolumn{3}{l}{$^{\mathrm{c}}$ from \citet{Geller2009} }\\
\hline
\end{tabular*}
\end{center}
\end{table*}

\section{OBSERVATIONS}
\label{sec:observations}
WOCS 5379 and WOCS 4540 were observed with COS on 2014 Nov 4 and 2014 Nov 10, respectively. Both targets were observed in TIME-TAG mode through the Primary Science Aperture (PSA) using the G140L grating with a central wavelength of 1105\AA. This region covers the Lyman-$\alpha$ wings while also providing the wide wavelength coverage necessary to fit WD atmosphere models in this region alone. Each target was observed for 6 orbits. The data were reduced through the MAST pipeline. To increase the signal-to-noise per resolution element, we binned every 30 resolution elements in each spectrum resulting in a final wavelength resolution of 2.4\AA.  

In both sources, the contribution from the BSS is seen in the red end of the spectrum. In order to isolate the WD emission, we exclude flux redward of 1750\AA\ for WOCS 5379 and 1625\AA\ for WOCS 4540. The BSS in WOCS 4540 is brighter and hotter than WOCS 5379 \citep[see Table~\ref{tab:binparams},][]{Gosnell2015} so the larger contamination is expected. We also mask out by hand geocoronal emission lines and one ISM absorption feature evident in the spectrum. The spectra are de-reddened assuming $E(B-V)=0.09$ \citep{Sarajedini1999}. 


\section{ANALYSIS}
\label{sec:analysis}

\subsection{Spectral Fitting with \texttt{emcee}}
\label{sec:fitting}

In order to determine the physical parameters of the WDs, the observed spectra are fit with model WD atmospheres using the python package \texttt{emcee}, an MCMC sampler \citep{emcee}. WD atmospheres are characterized by the effective temperature, \teff\ , and the surface gravity, \logg. Both observed spectra are fit using a grid of WD atmosphere models ranging from 11,000 K to 35,000 K in \teff\ and from 6.0 to 9.0 in \logg\ in steps of 0.25 \citep{Koester2010}. The models assume a thick H-atmosphere with $M_{H}/M_{tot}=10^{-4}$. Interpolation between the grid models is carried out with RegularGridInterpolator from \texttt{scipy} \citep{scipy}.

When fitting only the Lyman-$\alpha$ region, different flux normalizations result in fits of equal quality that are degenerate in \teff\ and \logg. Every pair of \teff\ and \logg\ values that are an appropriate fit to the Lyman-$\alpha$ region of a WD spectrum yields an associated radius given an adopted WD mass-radius relationship. Since the normalization of a WD model atmosphere to the observed spectrum is based on the ratio of the WD radius and the distance to the WD, $r^{2}/d^{2}$, the degeneracy can be broken if the distance to the WD is known. 

\subsubsection{Distance to NGC 188}
\label{sec:distance}

Whether the normalization of a WD atmosphere described by a particular pair of \teff\ and \logg\ values is accurate depends on how well the distance can be determined \citep{Landsman1996}.

Many distances to NGC 188 have been previously published. \citet{Sarajedini1999} found a distance of $1940\pm70$ pc from isochrone fitting. \citet{Meibom2009} used V12, an eclipsing binary near the turnoff of the cluster, to determine a distance of $1770\pm75$ pc. With the advent of the \textit{Gaia} mission, there are now parallax distances of individual cluster members \citep{gaiamission}. \citet{BailerJones2018} calculate individual distances for WOCS 4540 of $1980\pm70$ pc and WOCS 5379 of $1780\pm90$ using \textit{Gaia} DR2 data \citep{gaiadr2}. 

\begin{figure}
    \centering
    \includegraphics[scale=0.4]{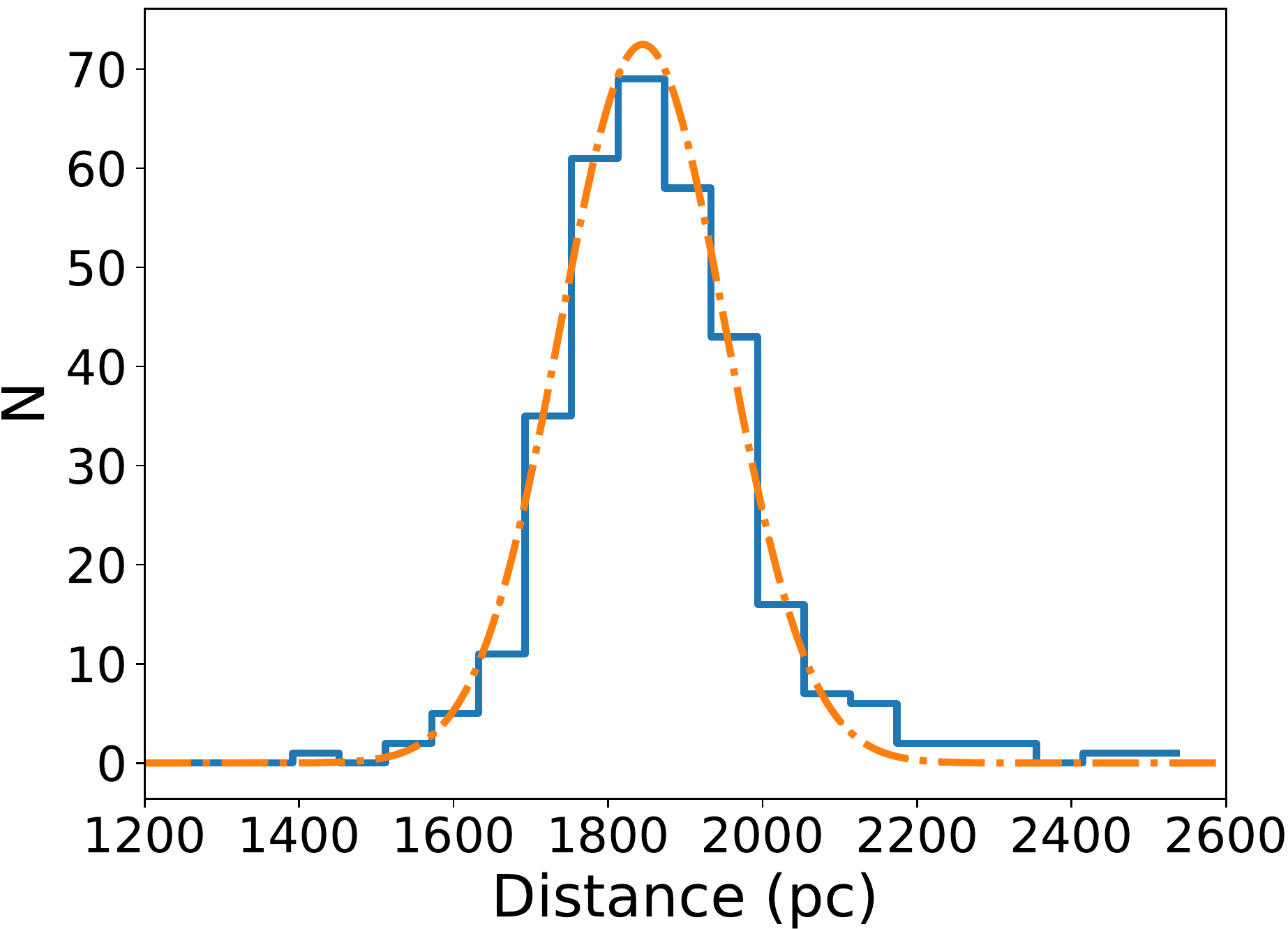}
    \caption{Histogram (blue) of distances from \citet{BailerJones2018} for 323 non-velocity variable cluster members of NGC 188 \citep{Geller2009} with low excess astrometric noise in \textit{Gaia} DR2 \citep{gaiadr2}. The distribution of distances is fit with a Gaussian to find a cluster distance of $1845\pm107$ pc, shown with the dot-dash orange line.}
    \label{fig:gaia_histogram}
\end{figure}

We also calculate a \textit{Gaia}-based cluster distance. We find the individual distances from \citet{BailerJones2018} for all ``Single Members" from the \citet{Geller2009} kinematic membership catalog for NGC 188, consisting of non-velocity variable stars that have high proper-motion and radial-velocity membership probabilities. We then remove sources that have excess astrometric noise in \textit{Gaia} DR2 greater than 0.001, resulting in a sample of 323 sources with distances between 1000 and 3000 pc. The resulting distribution of distances is shown in Figure~\ref{fig:gaia_histogram}. We fit a Gaussian distribution to this sample and find a distance to NGC 188 of $1845$ with a standard deviation of 107 pc. This is consistent with previously published \textit{Gaia}-based values for the distance to the center of the cluster \citep{Cantat-Gaudin2018}.

\subsubsection{Fitting Methodology}
\label{sec:metholodogy}

Although all the distances to NGC 188 presented in Section~\ref{sec:distance} are very similar, we fit both WDs using each distance to determine how the assumed distance impacts the fitted and derived WD parameters. For every instance of the fit, the distance for each system is independently sampled from a normal distribution with the mean and standard deviation described in Section~\ref{sec:distance} and listed in Table~\ref{tab:fitresults}. After the distance is selected, the normalization factor is calculated by finding the corresponding WD radius for a given pair of \teff\ and \logg\ values.

The mass-radius relationship, however, also depends on the core composition of the WD \citep{Holberg2006,Tremblay2011,Althaus2013}. Case B and Case C mass-transfer processes result in different WD core compositions, with Case B systems having He-core WDs and Case C systems having CO-core WDs. \citet{Parsons2017} find that all CO-core WDs in their sample of eclipsing WD binaries have masses above 0.50 \msun, while all He-core WDs have masses below 0.50 \msun. Although it is possible to create CO-core WDs with masses below 0.50 \msun\ \citep{Han2000,Willems2004,PradaMoroni2009}, the necessary scenarios require close binaries with a giant star mass of approximately 2.5 \msun. From previous photometric detection of the WDs in WOCS 4540 and WOCS 5379 \citep{Gosnell2014, Gosnell2015}, we know that these systems formed within the last 300 Myr when the turnoff mass of NGC 188 would only be 1.1--1.2 \msun. As this mass is well below the 2.5 \msun\ required to form an under-massive CO-core WD, we assume that this scenario is not possible for these mass-transfer systems in NGC 188. We do, however, allow for the possibility that a CO-core WD mass could go down to the approximate He-flash mass of 0.47 \msun\ for solar-metallicity stars \citep{Mocak2010,Bildsten2012}. We set a uniform \logg\ prior for a CO-core WD between 7.7 and 9.0, corresponding to an approximate minimum WD mass of 0.47 \msun\ for a moderate temperature WD \citep{Tremblay2011}. As He-core WDs are all expected to be below 0.47 \msun\, we set a uniform \logg\ prior for He-core fits to between 6.0 and 7.7 \citep{Althaus2013,Parsons2017}.

Both He-core and CO-core WDs can have thick H-atmospheres, resulting in spectra that qualitatively appear very similar. The possible surface gravity ranges are different, but since \teff\ also impacts the line shape of Ly-$\alpha$ the surface gravity alone is not a distinguishing characteristic. The different radii expected for CO-core WDs compared to He-core WDs, however, result in different atmosphere normalizations at the same distance. By constraining the distance and assuming a mass-radius relationship for both He-core WDs and CO-core WDs, the fit also constrains the range of physically realistic normalizations that are possible. In doing so, we determine whether either WD spectrum can be reasonably fit assuming a He-core WD or a CO-core WD.

Each system is fit twice, once using the \logg\ prior for He-core WDs and once using the \logg\ prior for CO-core WDs. For WOCS 4540, only the CO-core WD fit converges. Conversely, for WOCS 5379 only the He-core WD fit converges. Therefore, we conclude that WOCS 4540 is a CO-core WD and WOCS 5379 is a He-core WD. This is consistent with the types of WDs expected in each system given their different orbital periods \citep[see Table~\ref{tab:binparams},][]{Chen2008}. For the converging fits, \texttt{emcee} Markov chains are run with 1000 ``walkers'' each for 1000 samples. We then removed the burn-in, resulting in $8\times10^{5}$ samples of the posterior. 

The fit results for \teff\ and \logg\ for each distance assumption are provided in Table~\ref{tab:fitresults}, adopting the 16th and 84th percentile ranges as the uncertainties.

\begin{figure*}
\begin{center}
\includegraphics[scale=0.8]{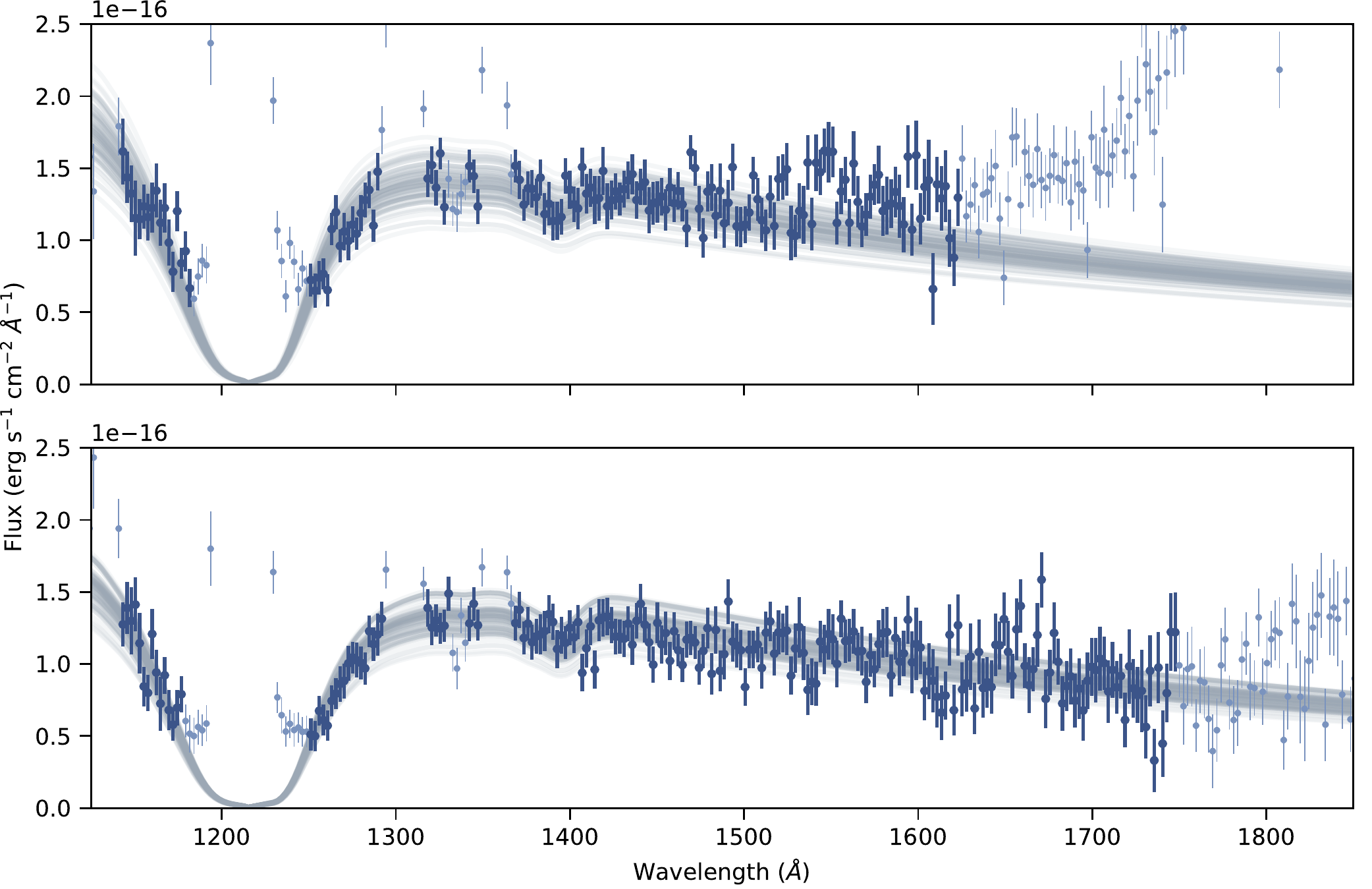}
\end{center}
\caption{CO-core WD atmosphere fits to the spectrum of WOCS 4540 (top) and He-core WD atmosphere fits WOCS 5379 (bottom). The data used in the fit are shown in darker blue, while the data masked from the fit are shown in light blue. The gray lines are 100 random draws from the posterior distribution, adopting the \textit{Gaia}-based cluster distance to NGC 188. The best fit parameters for WOCS 4540 are \teff\ $=17000^{+140}_{-200}$ K and \logg\ $=7.80^{+0.06}_{-0.06}$. The best fit parameters for WOCS 5379 are \teff\ $=15500^{+170}_{-150}$ and \logg\ $=7.50^{+0.06}_{-0.05}$. There is an excess in the spectrum for WOCS 4540 from approximately 1550 to 1625 \AA\ that is not well fit by the WD atmosphere and is likely additional contamination from the BSS primary.
}
\label{fig:modspec}
\end{figure*}

In Figure~\ref{fig:modspec} we plot 100 random draws from the posterior assuming the \textit{Gaia}-based cluster distance with each observed spectrum. The data used in the fit are shown in dark blue, with the masked sections of the spectrum (not used in the fit) shown in light blue. We note that the fit is not strongly dependent on the exact masking choices.

\begin{table*}
\begin{center}
\caption{Fitted and derived WD properties} \label{tab:fitresults}
\begin{tabular*}{6.6in}{@{\extracolsep{\fill}} c c c c c c c c c }
& \multicolumn{4}{c}{WOCS 4540} & \multicolumn{4}{c}{WOCS 5379} \\
Assumed  & \teff\ & \logg\ & Mass & Cooling & \teff\ & \logg\ & Mass & Cooling  \\
Distance (pc) & (K) & (cm s$^{-2}$) & ($M_\odot$) & age (Myr) & (K) & (cm s$^{-2}$) & ($M_\odot$) & age (Myr) \\
\hline
\vspace{4pt}
$1845 \pm 107^{\mathrm{a}}$ & 
$17000^{+140}_{-200}$ & $7.80^{+0.06}_{-0.06}$ & 
$0.53^{+0.03}_{-0.03} $ & $105^{+6}_{-5}$ &
$15500^{+170}_{-150}$  & $7.50^{+0.06}_{-0.05}$  & 
$0.42^{+0.02}_{-0.02}$ & $250^{+20}_{-20}$ \\
\vspace{4pt}
$1770\pm75^{\mathrm{b}}$ & 
$16900^{+160}_{-200}$ & $7.82^{+0.06}_{-0.05}$  &
$0.53^{+0.03}_{-0.03}$ & $110^{+5}_{-5}$ & 
$15500^{+200}_{-150}$ & $7.53^{+0.07}_{-0.05} $ &
$0.43^{+0.02}_{-0.02}$ & $260^{+20}_{-20}$ \\
\vspace{4pt}
$1940\pm70^{\mathrm{c}}$ & 
$17000^{+130}_{-140}$ & $7.78^{+0.04}_{-0.05} $ & 
$0.52^{+0.02}_{-0.02}$ & $100^{+5}_{-5}$ & 
$15500^{+160}_{-140}$ & $7.49^{+0.04}_{-0.05}$ & 
$0.42^{+0.02}_{-0.02}$ & $240^{+20}_{-20}$ \\

$1980\pm70$, $1780\pm90^{\mathrm{d}}$ & 
$17000^{+120}_{-120}$ & $7.77^{+0.04}_{-0.04}$ & 
$0.51^{+0.02}_{-0.02}$ & $100^{+5}_{-5}$ & 
$15500^{+150}_{-140}$ & $7.53^{+0.05}_{-0.05}$ &
$0.43^{+0.02}_{-0.02}$ & $260^{+15}_{-20}$ \\

\hline
\multicolumn{9}{l}{$^{\mathrm{a}}$ \textit{Gaia}-based cluster distance from this work, see Section~\ref{sec:distance} }\\
\multicolumn{9}{l}{$^{\mathrm{b}}$ Cluster distance from \citet{Meibom2009}   }\\
\multicolumn{9}{l}{$^{\mathrm{c}}$ Cluster distance from \citet{Sarajedini1999}  }\\
\multicolumn{9}{l}{$^{\mathrm{d}}$ Individual \textit{Gaia} DR2 distances from \citet{BailerJones2018} }\\
\end{tabular*}
\end{center}
\end{table*}

\subsection{Derived WD Properties}

For each distance assumption, we calculate the corresponding WD mass and age for each sample in the posterior by interpolating across the CO-core relationships from \citet{Holberg2006} and \citet{Tremblay2011} for WOCS 4540 (hereafter referred to as the ``Bergeron" grid\footnote{\url{http://www.astro.umontreal.ca/~bergeron/CoolingModels}}) and the He-core relationships from \citet{Althaus2013} for WOCS 5379. The derived values are given in Table~\ref{tab:fitresults} with the 16th to 84th percentile values given as uncertainties.

The fitted and derived properties of the WDs are very similar across all of the possible distances explored here, as all are consistent to one another within the 16th to 84th percentile ranges listed. For our remaining analyses, we adopt the results based on the \textit{Gaia}-based cluster distance of $1845\pm107$ pc (see Section~\ref{sec:distance}). At this distance WOCS 4540 is best fit with a CO-core WD with \teff\ $=17000^{+140}_{-200}$ K and \logg\ $=7.80^{+0.06}_{-0.06}$, corresponding to a WD mass of $0.53^{+0.03}_{-0.03}$ \msun\ and a cooling age of $105^{+6}_{-5}$ Myr. WOCS 5379 is best fit with a He-core WD with $=15500^{+170}_{-150}$ and \logg\ $=7.50^{+0.06}_{-0.05}$ corresponding to a WD mass of $0.42^{+0.02}_{-0.02}$ \msun\ and a cooling age of $250^{+20}_{-20}$ Myr. We provide the fitted and derived parameters at the other possible distances to both enable later studies wanting to adopt a different cluster distance and demonstrate the relationship between the distances and the WD properties. We note, though, that the masses and WD cooling ages at different distances are similar enough that our final analysis and interpretation of mass transfer scenarios in these systems is not strongly dependent on the adopted cluster distance.

\begin{figure*}[ht]
\begin{center}
\includegraphics[width=0.49\textwidth]{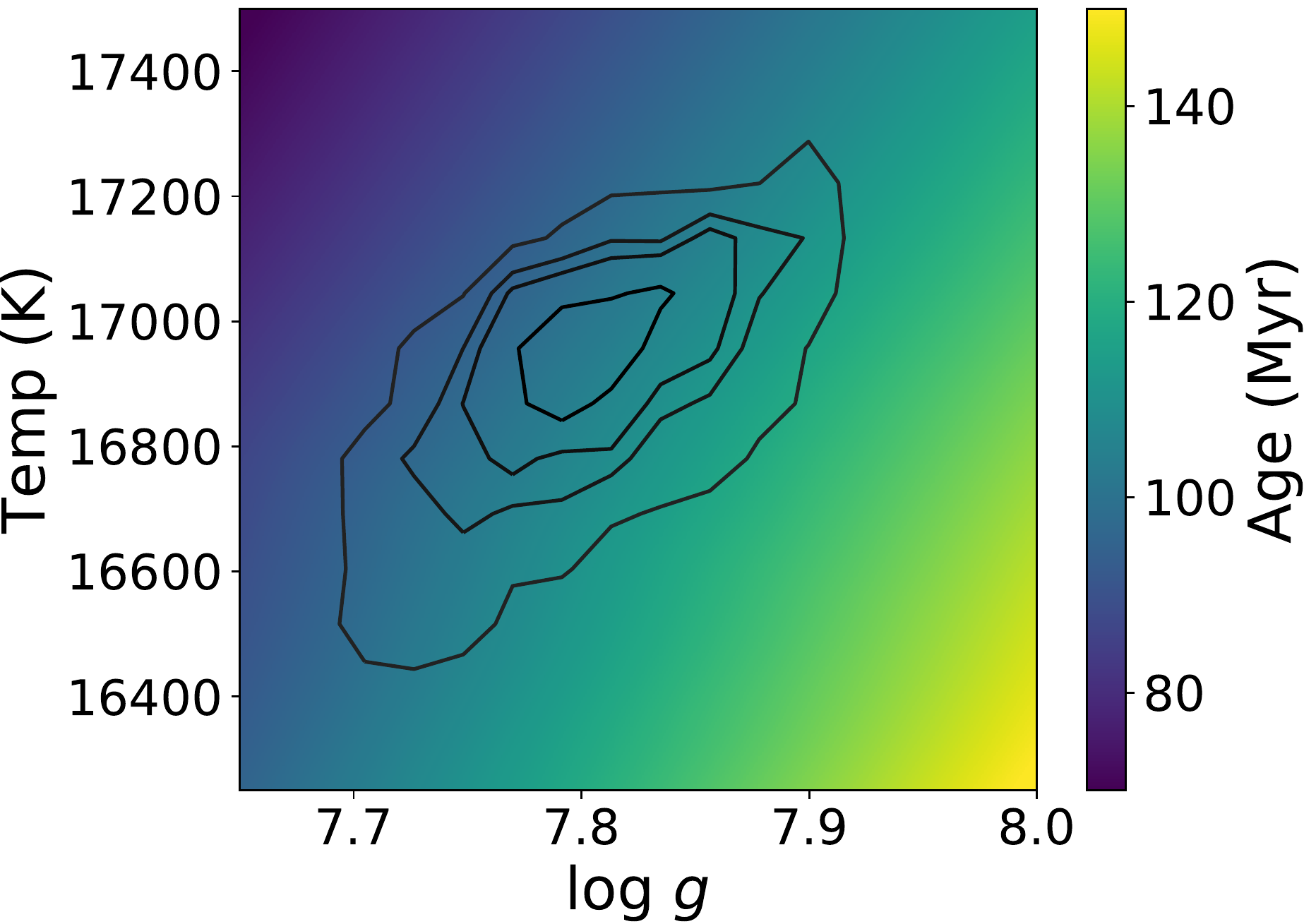}
\hspace{4mm}%
\includegraphics[width=0.47\textwidth]{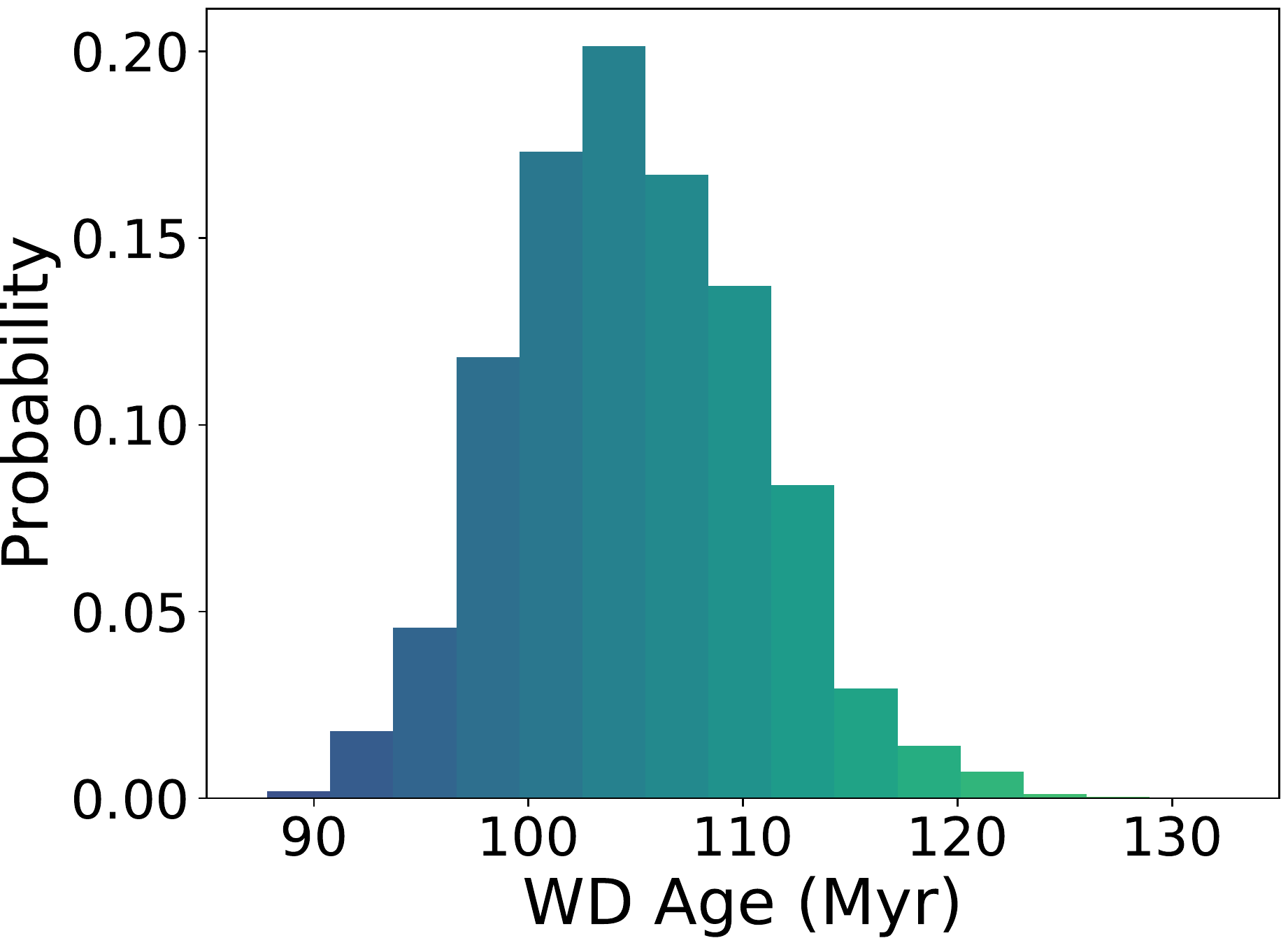}
\end{center}
\caption{On the left we show contours of the posterior distribution of the \texttt{emcee} fit for WOCS 4540 in \teff\ and \logg\ on top of a color map of the corresponding WD cooling ages through this range of \teff\ and \logg\ parameter space \citep{Holberg2006,Tremblay2011}, adopting the \textit{Gaia}-based cluster distance. The full color map is created from a bilinear interpolation of the original Bergeron grid. The corresponding histogram of ages for WOCS 4540 is shown on the right.
}
\label{fig:4540age}
\end{figure*}

\begin{figure*}[htbp]
\begin{center}
\includegraphics[width=0.49\textwidth]{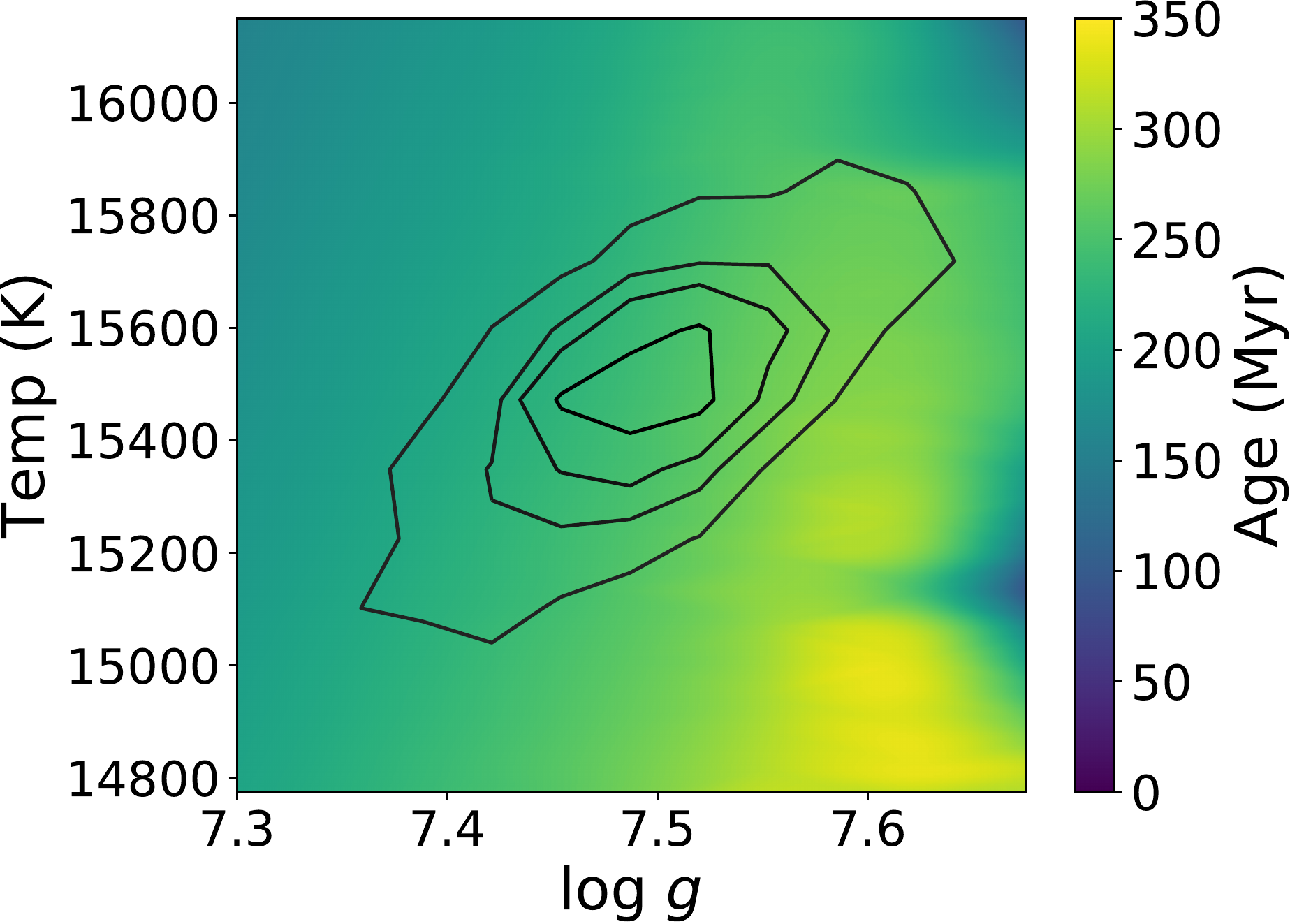}
\hspace{4mm}%
\includegraphics[width=0.47\textwidth]{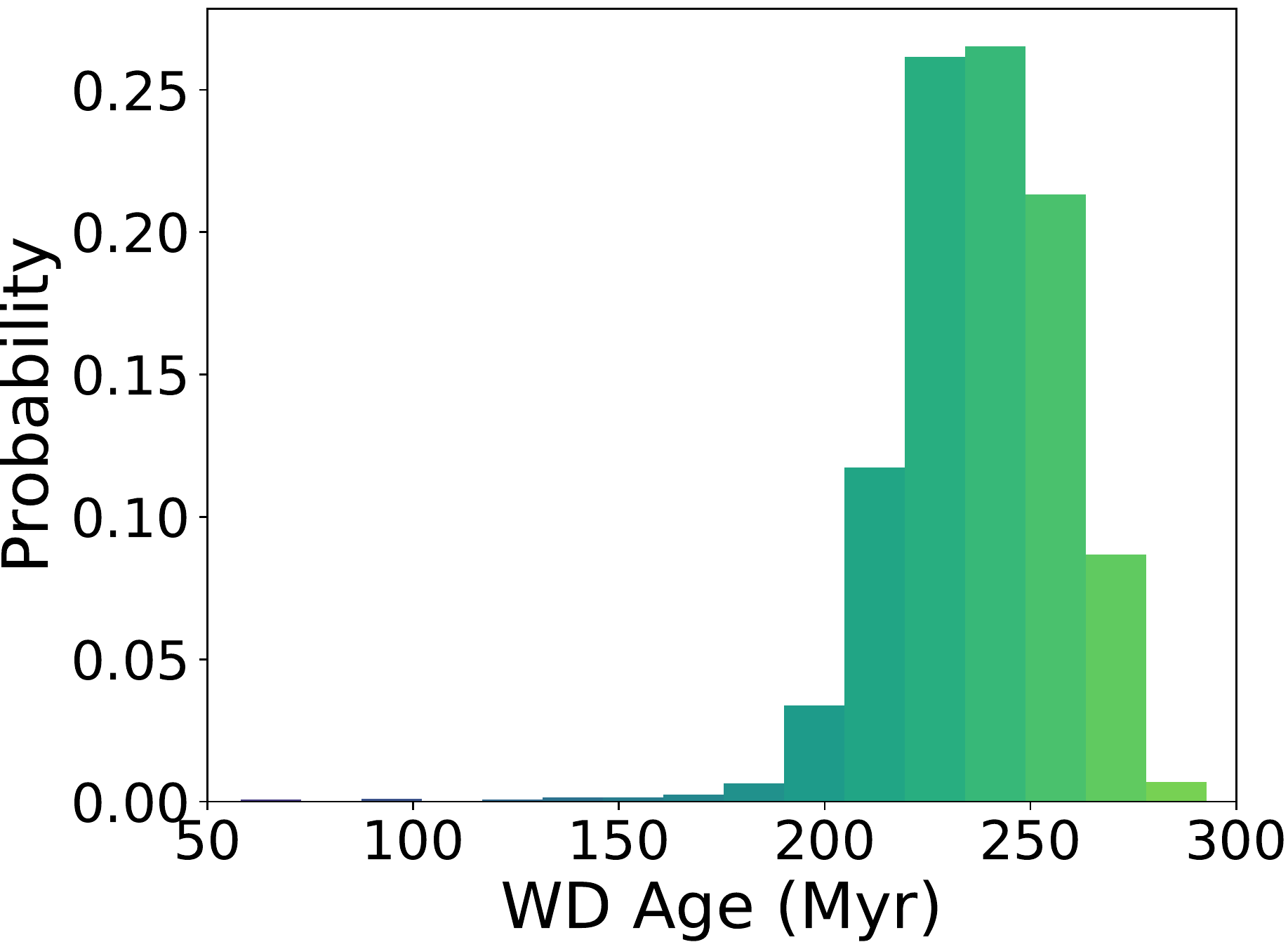}
\end{center}
\caption{Same as in Figure~\ref{fig:4540age}, but for WOCS 5379. Here the WD cooling ages are found using \citet{Althaus2013}. The structure apparent in the age color map results from the original \citet{Althaus2013} grid. In the histogram on the right, less than 1\% of the posterior exists in the tail toward low ages. }
\label{fig:5379age}
\end{figure*}

In Figures~\ref{fig:4540age} and~\ref{fig:5379age} we show contours of the posterior in \teff\ and \logg\ over a map of how these values correspond to the WD cooling age, and then the resulting histogram of ages from the posterior using a distance of $1845\pm107$ pc. The age color maps for WOCS 4540 and WOCS 5379 are created using a bilinear interpolation of the original grids of Bergeron and \citet{Althaus2013}, respectively. The age distribution for WOCS 5379 does have a tail toward younger ages, but the tail includes less than 1\% of the total posterior distribution. The peak and median of the distribution are both around 250 Myr.

\section{DISCUSSION}
\label{sec:discussion}
The presence of these systems in a well-studied cluster environment provides numerous constraints on the pre- and post-mass-transfer binary parameters. Adopting a distance to NGC 188 of 1845 pc (Section~\ref{sec:distance}) and a reddening of $E(B-V)=0.09$ \citep{Sarajedini1999}, we use Modules for Experiments in Stellar Astrophysics \citep[MESA, version 8118,][]{mesa} to establish a 6.2 Gyr age and a turnoff mass of 1.1 \msun, which is consistent with the age from \citet{Meibom2009}. Between 100--300 Myr ago, the main-sequence turnoff was approximately 1.2 \msun. We assume that both WOCS 4540 and WOCS 5379 originally had a 1.2 \msun\ primary, and hypothesize that differing progenitor binary parameters resulted in two very different BSS systems. 

\begin{figure}[h]
\begin{center}
\includegraphics[scale=0.33]{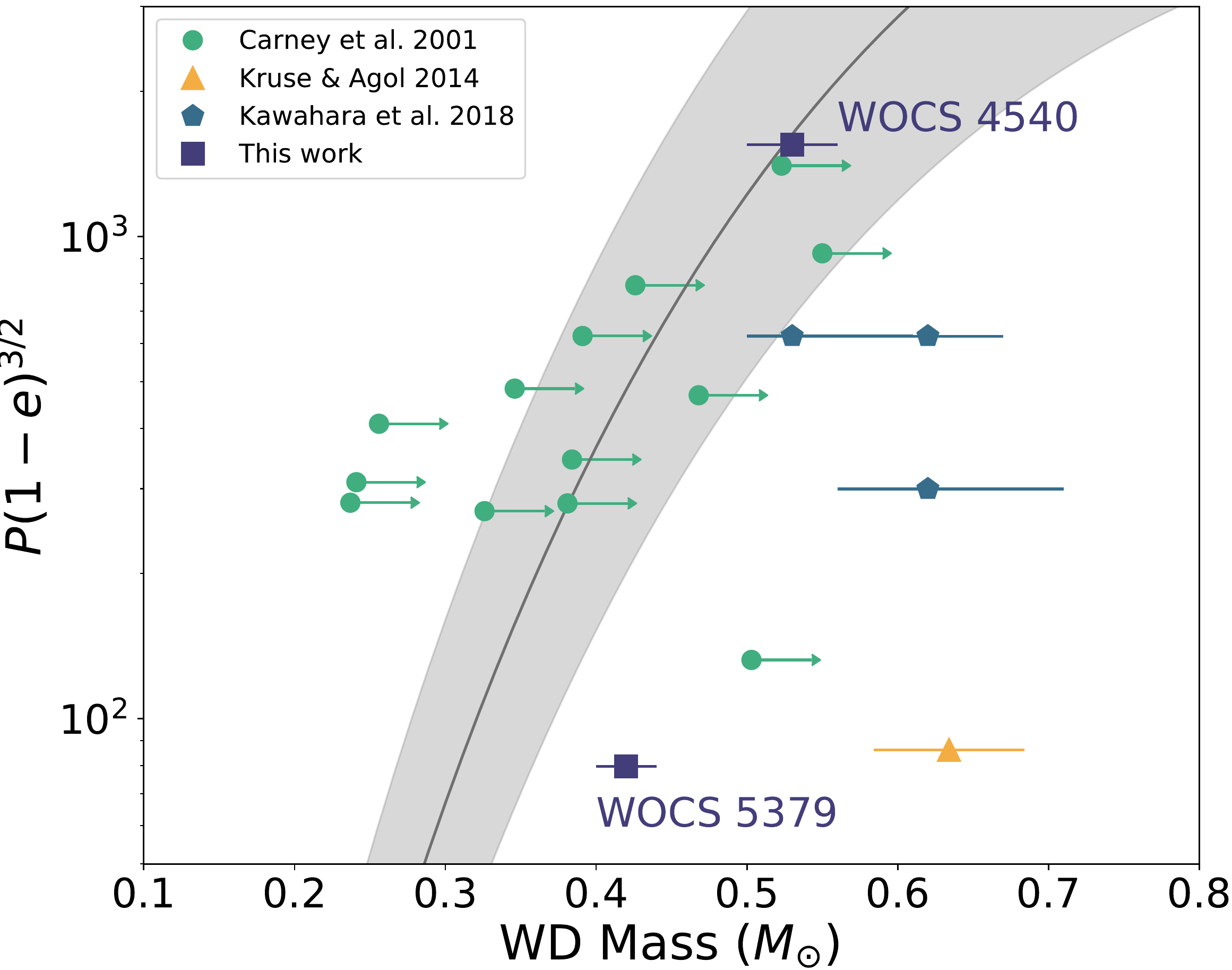}
\end{center}
\caption{Post-mass transfer WD binaries in the $P$--$M_{WD}$ plane, modeled after \citet{Kawahara2018}. In addition to the two binaries in this study, we include the sample of field BSS binaries from \citet{Carney2001} and the self-lensing WD binaries from \citet{Kruse2014} and \citet{Kawahara2018}. As many BSS binaries are eccentric, we plot WD mass against the instantaneous period at periastron, $P(1-e)^{3/2}$, instead of just against the orbital period. In gray we show the \citet{Rappaport1995} theoretical prediction for stable mass transfer with the upper and lower limits of the relationship (factors of 2.4) in the gray shaded region. Although some post-mass-transfer systems are consistent with the prediction for stable mass transfer, including WOCS 4540, there are numerous shorter-period systems that fall below the expectations for stable mass transfer, such as WOCS 5379. The error bars for WOCS 4540 and WOCS 5279 represent the 16th to 84th percentile range for the mass after adopting the \textit{Gaia}-based cluster distance.} 
\label{fig:massperiod}
\end{figure}

We first compare these systems to the \citet{Rappaport1995} theoretical relationship between final orbital period and WD mass for systems created through stable mass transfer, as shown in Figure~\ref{fig:massperiod}. Many BSS binaries are eccentric, so we plot the WD mass against the  instantaneous period at periastron ($P(1-e)^{3/2}$), rather than the orbital period. The error bars for WOCS 4540 and WOCS 5379 show the 16th to 84th percentile range adopting the \textit{Gaia}-based cluster distance, although the mass range for each source is very similar across all the distances in Table~\ref{tab:fitresults}. 

To illustrate the diversity of post-mass transfer binaries we also include the sample of field BSSs from \citet{Carney2001} and self-lensing WD binary systems from \citet{Kawahara2018} and \citet{Kruse2014}. In general, many systems are consistent with the \citet{Rappaport1995} theoretical prediction for stable mass transfer, including WOCS 4540. There are several systems that fall below this relationship, however, suggesting that post-mass-transfer products can form through non-stable mass-transfer processes, including WOCS 5379. We return to this topic in Sections~\ref{sec:4540} and~\ref{sec:5379}.

\subsection{MESA mass accretion models}
\label{sec:accretionmodels}
To explore the range of possible progenitor binaries we run a grid of MESA mass accretion models \citep{mesa}. Comparing the HR Diagram positions of WOCS 4540 and WOCS 5379 to normal MESA evolutionary models suggests that the current observed masses of WOCS 4540 and WOCS 5379 are around 1.5--1.6 \msun\ and 1.2--1.3 \msun, respectively, but we know that these stars do not have standard evolutionary histories. We instead add mass to a main-sequence star such that it matches the observed characteristics of these systems at the 6.2 Gyr age of NGC 188. We are only interested in the behavior of the accreting star (that will become the BSS), so we do not model the full binary evolution (which is beyond the scope of this paper). All models are single-star models; we do not consider the properties of the donor or the orbital parameters of the binary system. We use the MESA test suite case  \texttt{1M\char`_pre\char`_ms\char`_to\char`_wd}, but we turn off stellar rotation and do not consider rotational mixing. We assume no wind mass loss, since we are only interested in the main-sequence evolution of the accretor, where the wind mass loss rates should be small. 

We evolve progenitor systems ranging from 0.7--1.1 \msun\ in increments of 0.1 \msun\ up to an age of approximately 6 Gyr. At this point, the star accretes mass onto its surface until it reaches either 1.5 or 1.6 \msun\ for WOCS 4540 or 1.2 or 1.3 \msun\ for WOCS 5379. If the mass transfer is fully conservative, all of the mass leaving the donor will be accreted by the proto-BSS, and the accretion rate should equal the donor mass loss rate. Assuming the donor loses the entire envelope during the extent of the RGB or AGB phase the mass loss rate would be $10^{-8}$ to $10^{-6}$ \msun\ yr$^{-1}$, respectively. If the mass transfer is not conservative the accretion rate will be less than the donor mass-loss rate. 

As we are not carrying out actual binary evolution modeling, we set the accretion rate at a constant value. We adopt a rate of $1 \times 10^{-8}$ \msun~yr$^{-1}$ which keeps the accreting stars in thermal equilibrium during accretion. We note, however, that the choice of mass-transfer rate has very little impact on the final properties of the modeled BSSs. The WD cooling ages for both systems are longer than the estimated thermal timescale of 70 Myr for both WOCS 4540 and WOCS 5379. If either system is the result of a higher mass-transfer rate that drove the accretor out of thermal equilibrium, the BSS has had enough time to readjust. For this reason, we also assume that the final HR diagram position of the BSS does not strongly depend on whether the accretion rate was constant or variable, so our choice of a constant accretion rate is reasonable.

Adopting 6.2 Gyr as the current age for NGC 188, we time the start of mass accretion such that it ceases at an age appropriate to the cooling age of the WD companions. We assume the composition of the accreted material matches the surface composition of the accreting star. We then allow the BSS to continue evolving to the giant branch, terminating the models when the helium core mass reaches 0.2 \msun. We note that the models do not take into account any internal mixing or possible non-standard abundances as a result of the mass transfer history (see Section~\ref{sec:5379} for further discussion).

\begin{figure*}[ht]
\begin{center}
\includegraphics[scale=0.4]{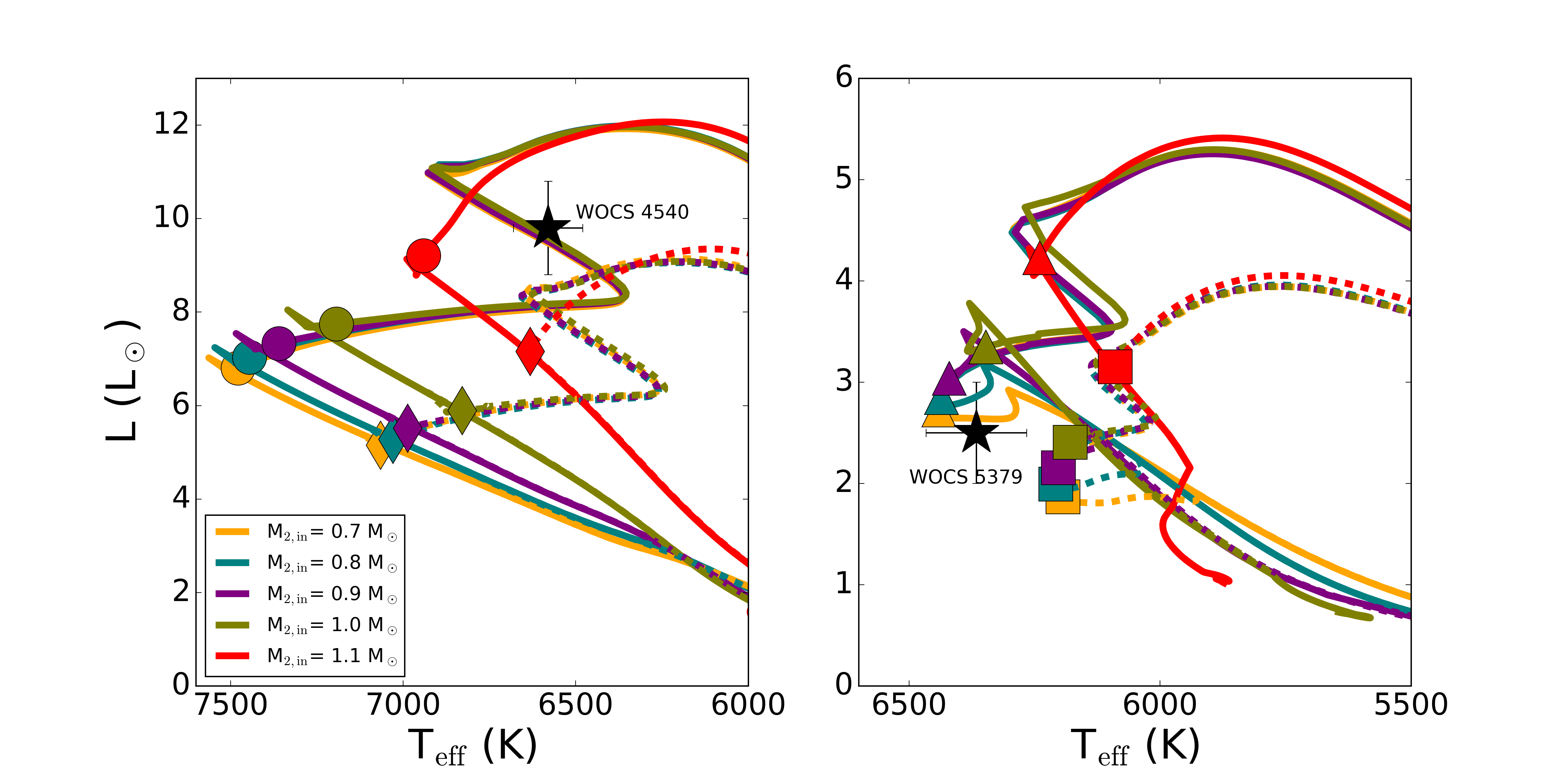}
\end{center}
\caption{MESA evolutionary tracks for accreting stars with initial masses ranging from 0.7 to 1.1 \msun, as shown in the figure legend. On the left, the tracks are compared against the \teff\ and luminosity of WOCS 4540, shown with a black star. Dashed lines show the accretion tracks ending with a 1.5 \msun\ BSS and the solid lines show tracks ending with a 1.6 \msun\ BSS, with the color indicating the original progenitor mass. The diamonds and circles show the locations on those tracks corresponding to an age of 6.2 Gyr. On the right, similar tracks are shown for WOCS 5379, with the dashed lines indicating a 1.2 \msun\ BSS and the solid lines indicating a 1.3 \msun\ BSS. The squares and triangles show the locations at 6.2 Gyr. These tracks show that the progenitor for WOCS 4540 must be slightly more massive than the assumed current turnoff mass of 1.1 \msun\ and the resulting BSS mass is between 1.5 and 1.6 \msun. WOCS 5379 has a progenitor of approximately 0.7 \msun\ with a final BSS mass close to 1.3 \msun.}
\label{fig:accretiontracks}
\end{figure*}

To compare the models to the BSSs we calculate the bolometric luminosities of WOCS 4540 and WOCS 5379 using the relationships in \citet{Torres2010} with temperatures from \citet{Gosnell2015}, resulting in a luminosity of $9.8\pm0.9$ $L_{\odot}$ for WOCS 4540 and $2.5\pm0.5$ $L_{\odot}$ for WOCS 5379. WOCS 5379 is a photometric variable \citep{Kafka2003}, so the luminosity uncertainty includes the 0.22 mag variation in the $V$-band. The resulting evolution tracks are shown in HR diagrams in Figure~\ref{fig:accretiontracks}. 

The symbols in Figure~\ref{fig:accretiontracks} show the location on each accretion track corresponding to an age of 6.2 Gyr, the current age of NGC 188. Comparing these symbols to the observed properties of the BSSs reveals the progenitor masses and final BSS masses most consistent with our models. WOCS 4540 appears to be between 1.5 and 1.6 \msun, requiring an initial progenitor of just over 1.1 \msun\ which is just beyond the grid of models used here. (Note that stars with progenitor masses of 1.2 \msun\ and higher have already evolved to the giant branch.) WOCS 5379 is close to a 1.3 \msun\ BSS forming from a 0.7 \msun\ progenitor. Due to the large luminosity uncertainty for WOCS 5379, though, it could have a progenitor as massive as 0.9 \msun.

\begin{figure*}[ht]
\begin{center}
\includegraphics[scale=0.4]{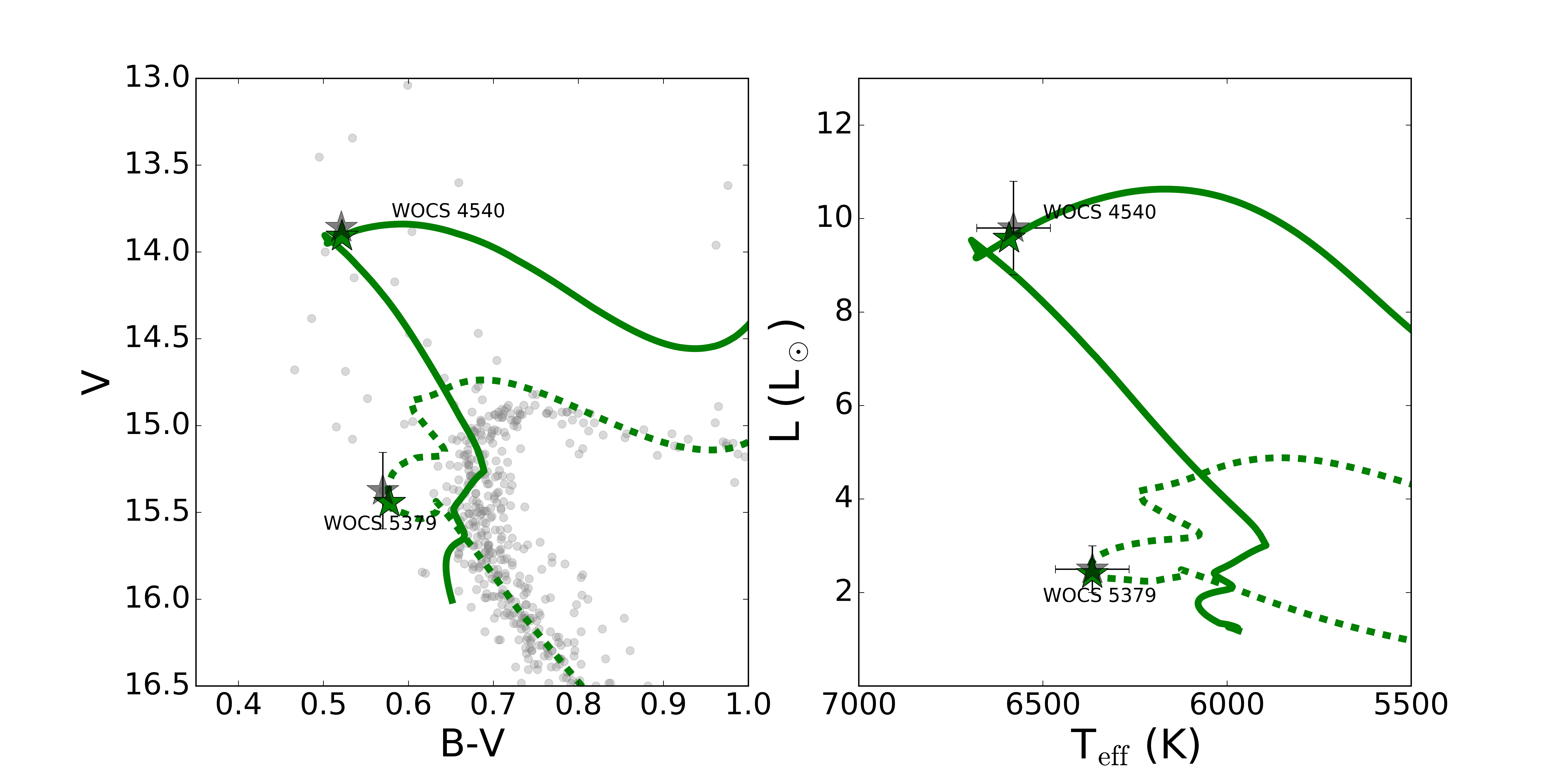}
\end{center}
\caption{Left: Color-magnitude diagram showing MESA accretion evolutionary tracks for both WOCS 4540 (solid line) and WOCS 5379 (dashed line) that best match the observed photometry. The gray dots show the NGC 188 cluster members \citep{Geller2009}. The current CMD positions of both systems are shown with gray stars. The photometric errors are within the points, but WOCS 5379 is a photometric variable with $\Delta$V$= 0.22$, which we show with the error bar. The modeled photometry for each track at an age of 6.2 Gyr is shown with green stars. The model that best matches WOCS 4540 is a 1.14 \msun\ progenitor that accretes 0.41 \msun\ to reach a final BSS mass of 1.55 \msun. The model that best matches WOCS 5379 is a 0.70 \msun\ progenitor that accretes 0.57 \msun\ to reach a BSS mass of 1.27 \msun. Right: The same models as on the left, but shown on an HR diagram.\label{fig:cmd}}
\label{fig:bestfittracks}
\end{figure*}

In addition to comparing model temperatures and luminosities to these BSSs, we use the MESA colors module to compute colors and magnitudes and find models that best match the observed photometry of these systems. Here we use the spectral library of \citet{Lejeune1998} to compute synthetic, un-extincted $B$ and $V$ magnitudes for these stars. We then use a reddening of $E(B-V)= 0.09$ \citep{Sarajedini1999} and assume a distance of 1847 pc to adjust these models to the distance and reddening of NGC 188. To find the model that best matches the photometry we explore a finer grid of MESA models for WOCS 4540 with progenitors of 1.1--1.15 \msun\ and adopt a progenitor mass of 0.7 \msun\ for WOCS 5379. The models that provide the closest match to the observed BSS photometry correspond to a a 1.14~\msun\ progenitor accreting mass up to 1.55 \msun\ for WOCS 4540, and a 0.7 \msun~progenitor accreting mass up to 1.27 \msun\ for WOCS 5379. These models are plotted on a CMD and HR diagram of NGC 188 in Figure~\ref{fig:bestfittracks}. 

The final positions of the modeled stars in Figures~\ref{fig:accretiontracks} and~\ref{fig:bestfittracks} are essentially the same if we instead assume a mass transfer rate of $1 \times 10^{-6}$ \msun~yr$^{-1}$, but the accretion paths go out of thermal equilibrium before relaxing to a similar luminosity and temperature by 6.2 Gyr. These accretion tracks only take into account the BSSs and do not include the stability of the mass transfer or the mass transfer efficiencies required, which are explored in the next Section.

A sample inlist and \texttt{run\_star\_extras.f} routine for this model grid can be found on \url{mesastar.org}.

\subsection{Mass transfer stability}
\label{sec:stability}
Comparing accretion tracks to the observed properties of the BSSs constrains the parameter space of possible progenitor systems, but it does not investigate the physical response of the donor star to mass transfer, which will likely determine whether the mass transfer is expected to be stable or unstable. Additionally, the accretion tracks alone also do not take into account mass-transfer efficiency, which is dependent on assumptions made of the donor star wind mass loss. 

To understand if these accretion scenarios are possible in the context of stable Roche lobe overflow and the state of the donor star, we compute the predicted stability of the model tracks described in Section \ref{sec:accretionmodels}. To do this, we compare the predicted adiabatic response of an evolved star to mass loss to the Roche lobe response to mass loss, given various binary mass ratios and mass transfer efficiencies. For example, if the Roche lobe responds to mass loss by shrinking and the giant star responds by expanding, we would expect runaway, unstable mass transfer to occur. If, on the other hand, we expect the giant to expand while the Roche lobe also expands we could expect stable mass transfer to occur. We make these calculations at the point when mass transfer begins. We assume that mass transfer that begins as stable or unstable due to the adiabatic response of the giant star would remain so throughout the mass-transfer process. Expressions for the Roche lobe and adiabatic response factors assuming a polytropic model are given in \citet{Ivanova2015} (see also \citealt{Hjellming1987}). 

The Roche lobe response factor is given by: 
\begin{equation}
\zeta_{RL}= \frac{\partial{~\mathrm{ln} a}}{\partial{~\mathrm{ln}m_d}}+\frac{\partial{~\mathrm{ln}(R_{RL}/a)}}{\partial{~\mathrm{ln} q}}\frac{\partial~\mathrm{ln}q}{\partial~\mathrm{ln}m_d}
\end{equation}
where  $R_{RL}$ is the Roche lobe radius, $a$ is the binary orbital separation, $m_d$ is the donor mass, and $q$ is the mass ratio of the system ($q= M_\mathrm{donor}/M_\mathrm{accretor}$). Using the Roche lobe approximation of \citet{Eggleton1983}, these partial differential equations can be approximated as: 
\begin{equation}
\frac{\partial~{\mathrm{ln} a}}{\partial{~\mathrm{ln}m_d}}= \frac{2q^2-2-q(1-B)}{q+1}
\end{equation}

\begin{equation}
\frac{\partial~\mathrm{ln}q}{\partial~\mathrm{ln}m_d}= 1+ B q
\end{equation}

\begin{equation}
\frac{\partial{~\mathrm{ln}(R_{RL}/a)}}{\partial{~\mathrm{ln} q}}= \frac{2}{3}- \bigg(\frac{q^{\frac{1}{3}}}{3}\bigg)\frac{1.2q^\frac{1}{3}+ 1/(1+q^\frac{1}{3})}{0.6q^\frac{2}{3}+\mathrm{ln}(1+q^\frac{1}{3})}
\end{equation}
where $B$ is the mass transfer efficiency (i.e., the fraction of the total mass lost by the donor that is accreted by the accretor).

The giant's adiabatic response to mass loss is given by: 

\begin{equation}
\begin{split}
\zeta_{ad} &= \frac{2}{3}\bigg(\frac{m}{1-m}\bigg)-\frac{1}{3}\bigg(\frac{1-m}{1+2m}\bigg) \\
&\phantom{E}-0.03+0.2\bigg[\frac{m}{1+(1-m)^{-6}}\bigg]
\end{split}
\end{equation}

\noindent where m is the ratio of the donor's core mass to total mass, $m= M_\mathrm{core}/M_\mathrm{donor}$.

We calculate $\zeta_{RL}$ from Equations 1--4 for every MESA model in Figure~\ref{fig:accretiontracks}. Here we use two different assumptions about the donor mass (taken either with or without wind mass loss, as described below) with the initial secondary mass of each model track adopted as the mass of the accreting star. The average mass transfer efficiency is calculated from: 
\begin{equation} \label{eq:efficiency}
B= \frac{M_\mathrm{2, f}- M_\mathrm{2, in}}{M_d- M_\mathrm{WD}}
\end{equation}
where $M_\mathrm{2, f}$ is the final mass of the modeled BSS from our grid, $M_\mathrm{2, in}$ is the initial mass of the BSS progenitor, $M_d$ is the donor mass, and $M_\mathrm{WD}$ is the final white dwarf mass (which we take to be either 0.42 \msun\ or 0.53 \msun). 

\begin{figure*}
\begin{center}
\includegraphics[scale=0.4]{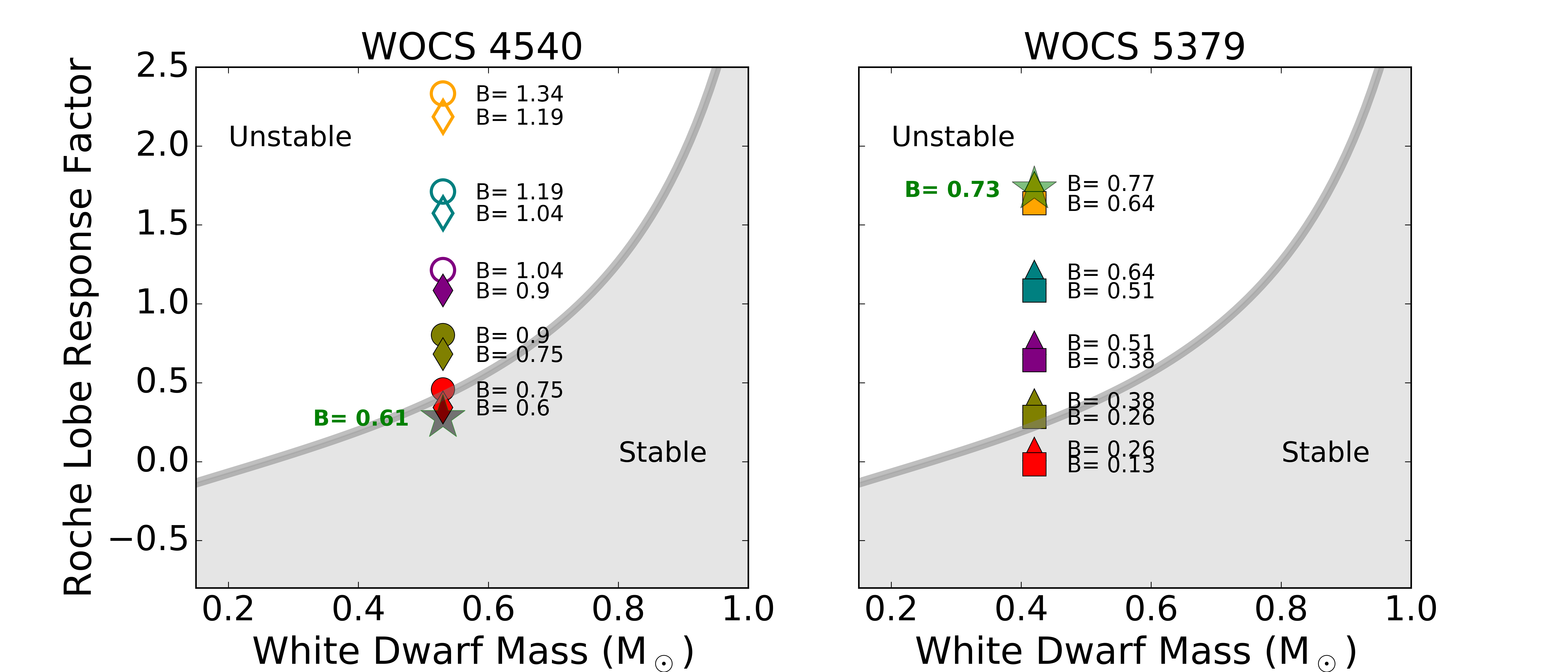}
\end{center}
\caption{Stability calculations for the formation models shown in Figures \ref{fig:accretiontracks} and \ref{fig:cmd}, assuming the donor star undergoes no wind mass loss prior to the onset of Roche lobe overflow. Models for WOCS 4540 are shown on the left, and models for WOCS 5379 are on the right. Plots compare the response of the Roche lobe to mass loss with the expected adiabatic response of the giants to mass loss as a function of the resulting white dwarf mass to the formation models (colored symbols). The shaded gray region represents mass transfer scenarios expected to be stable, where $\zeta_{RL} > \zeta_{ad}$, and the white region indicates the unstable mass transfer regime ($\zeta_{RL} < \zeta_{ad}$) where the mass-losing giant is expected to expand faster than its Roche lobe. The grey line is where $\zeta_{RL} = \zeta_{ad}$. Symbols correspond to different final blue straggler masses (1.2--1.6 \msun), and colors correspond to initial accretor masses (0.7--1.1 \msun) as in Figure \ref{fig:accretiontracks}. The green stars indicate the stability of the best-fit evolutionary tracks from Figure~\ref{fig:cmd}. For each model we also show the necessary mass transfer efficiency (B). Given the assumptions made about the donor star, some of the models require mass transfer efficiencies greater than 1, meaning the BSSs would require more mass to be accreted than is available from the donor. These models are not physically reasonable, and are indicated with open symbols.}
\label{fig:stability}
\end{figure*}

\begin{figure*}
\begin{center}
\includegraphics[scale=0.4]{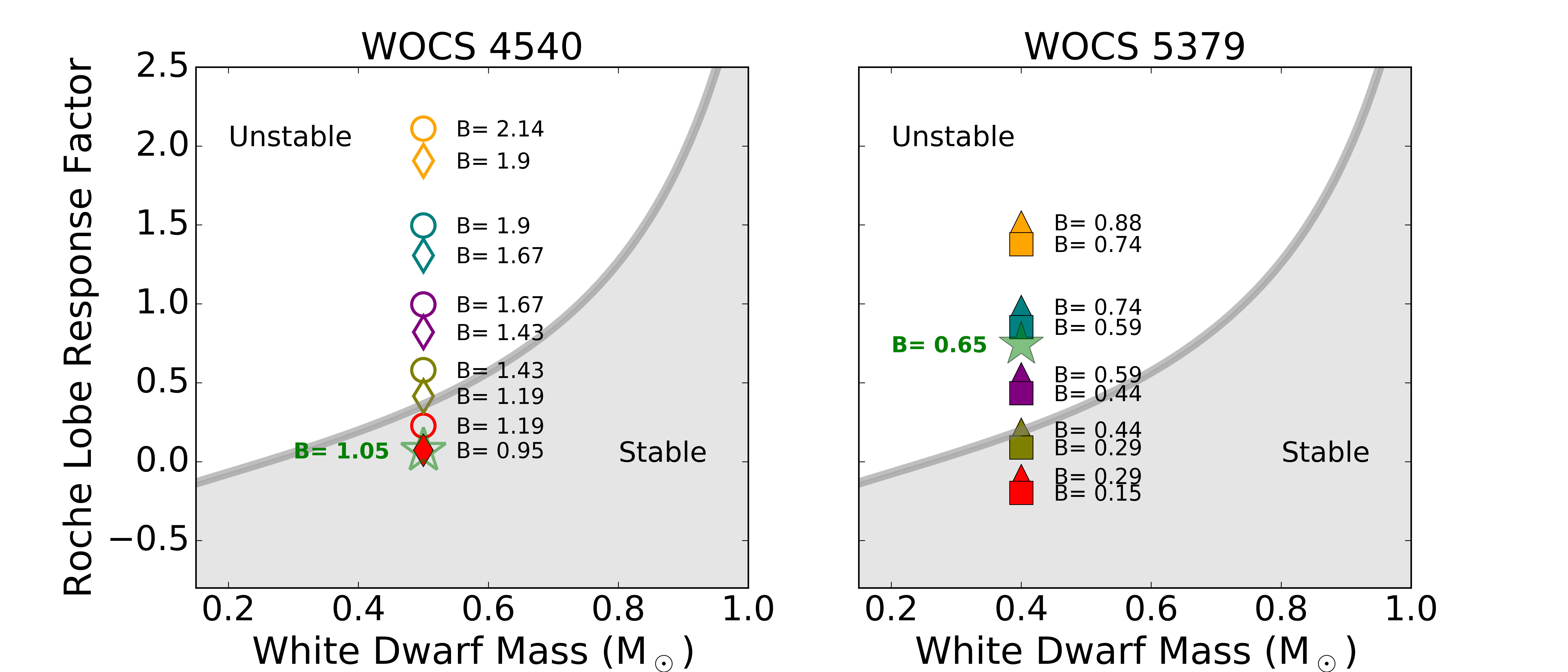}
\end{center}
\caption{Stability calculations as in Figure \ref{fig:stability}, but here we determine the adiabatic response of the giant donor assuming substantial wind mass loss prior to the onset of Roche lobe overflow. When accretion begins, the AGB donor for WOCS 4540 (left) has a mass of  0.9 \msun. The RGB donor for WOCS 5379 (right) has a mass of 1.1 \msun. We show the mass transfer efficiency (B) as a fraction of the mass lost during Roche lobe overflow (i.e. assuming none of the material lost earlier as a wind can be accreted by the secondary), with scenarios requiring a mass transfer efficiency above 1 marked with open symbols. This scenario provides an optimistic limit on the possible stability of these formation scenarios.}
\label{fig:stabilitywithwind}
\end{figure*}

We calculate $\zeta_{ad}$ from Equation 5 using two different sets of assumptions about the donor star. In Figures~\ref{fig:stability} and~\ref{fig:stabilitywithwind} we compare $\zeta_{RL}$ (solid gray line) to $\zeta_{ad}$ for each MESA model in Figure~\ref{fig:accretiontracks}.  Where $\zeta_{RL} >  \zeta_{ad}$ (gray shaded region), we expect Roche lobe overflow to lead to stable mass transfer. Where  $\zeta_{RL} < \zeta_{ad}$ (white region) we expect unstable mass transfer. Therefore, any models (colored points) that fall within the gray shaded region can plausibly result from stable Roche lobe overflow. In models above the gray shaded region we would expect Roche lobe overflow to be unstable.

In Figure \ref{fig:stability} we use $M_\mathrm{donor}= 1.2$ \msun, the expected mass of an NGC 188 RGB or AGB star at an approximate age of 6 Gyr given no wind mass loss. Next to each point on Figure~\ref{fig:stability} we show the overall mass-transfer efficiency ($B$). Note that for some of our models the mass-transfer efficiency is greater than 1, indicated with an open symbol. This indicates a case in which the main-sequence accretor must gain more material than is available from the donor in order to reach its final mass. This is nonphysical, indicating these mass-transfer models could not have occurred given our derived masses for the components of these systems. 

Additionally, unstable mass transfer would likely lead to a common-envelope event, which is thought to have a very low mass-transfer efficiency of less than 10\% \citep{Woods2011}. In contrast, our best-fit models in Figure~\ref{fig:stability} (green stars) require mass-transfer efficiencies over 60\%. This suggests that even unstable common-envelope evolution is unrealistic for these systems. We return to this topic in Section~\ref{sec:5379}.

We also consider the effect of wind mass loss on the giant branch on the stability of mass transfer, as shown in Figure~\ref{fig:stabilitywithwind}. Here we follow the same formalism as above, but the donor-star masses are determined after including a Reimers wind model \citep{Reimers1975} on the RGB and a Bloecker wind on the AGB \citep{Bloecker1995} with a scaling factor of $\eta=0.7$. Using this wind scheme, an RGB star with an initial mass of 1.2 \msun~will have a total mass of approximately 1.1 \msun\ in the middle of the RGB, and a total mass of approximately 0.9 \msun\ as it begins to ascend the AGB. Assuming these values as donor masses for WOCS 5379 and 4540 respectively, we recalculate the mass-transfer efficiency and stability shown in Fig. ~\ref{fig:stability}. This allows us to consider what may happen if a significant amount of wind mass loss occurs prior to Roche lobe overflow. 

In general, reducing the donor star mass through wind mass loss moves the models toward greater stability, but increases the mass transfer efficiency needed during Roche lobe overflow. Wind mass loss rates on the RGB and AGB are uncertain, but we set the scaling factor at the upper end of what has been calibrated in globular clusters \citep{McDonald2015}. This provides a limit on the mass-transfer scenarios that are potentially stable. We consider that a model that is unstable even with this upper-limit wind mass loss rate is unlikely to be stable when using RGB wind schemes typically found in the literature, resulting in more massive giant stars than assumed in Figure~\ref{fig:stabilitywithwind}.

Many of the accretion scenarios modeled in Figure~\ref{fig:accretiontracks} for WOCS 4540 are not likely, either because there is not enough mass to produce the necessary BSS mass or because the mass transfer would be unstable given the secondary mass and required mass-transfer efficiency. Only models with $B < 1$ that also fall within the gray stability region of Figures \ref{fig:stability} and \ref{fig:stabilitywithwind} are physically plausible, leaving only the red tracks as viable. However, the model that best matches the photometry (green star) is found to be stable regardless of the wind mass loss parameters. Although the maximal wind loss scenario that best matches the photometry requires more than 100\% mass transfer efficiency, the true wind loss of the system is likely lower and would have a total mass-transfer efficiency below 100\%. Additionally, the total mass-transfer efficiency needed is reduced if some wind mass transfer occurs in addition to stable mass transfer, which we return to in Section~\ref{sec:4540}. We therefore conclude that WOCS 4540 can plausibly form via a standard stable mass transfer process. 

On the other hand, even when using the more stable models that incorporate wind mass loss (Figure~\ref{fig:stabilitywithwind}), the models that best match the photometry of WOCS 5379 (green star) is still unstable. WOCS 5379 is therefore an intriguing system that challenges our assumptions about how and when mass transfer occurs, which we return to in Section~\ref{sec:5379}.

\subsection{Formation Pathway for WOCS 4540}
\label{sec:4540}

The current period of WOCS 4540 of $3030\pm70$ days is traditionally thought to be too wide for Case C mass transfer \citep{Chen2008}. The eccentricity of the binary, however, brings the periastron separation of the stars close enough to be consistent with theoretical predictions for stable mass transfer, as shown in Figure~\ref{fig:massperiod}. Although traditional binary interaction models assume that all mass transfer binaries are circularized, mass transfer from an AGB star at periastron passage can combat tidal circularization, resulting in an eccentric post-mass-transfer system \citep{Soker2000,Bonacic2008}. 

Additionally, we show in Section~\ref{sec:stability} that the accretion evolutionary tracks that can reasonably describe the current HR Diagram and CMD location of WOCS 4540 are also consistent with a stable mass-transfer history. 

We also consider the possibility that WOCS 4540 formed through an efficient wind accretion mechanism, such as Wind Roche Lobe Overflow (WRLOF) \citep{Mohamed2007,Abate2013}. This mechanism requires the AGB wind dust-formation radius to be beyond the Roche lobe radius of the system. If the wind within the dust-formation radius has a velocity less than the escape velocity of the system, the Roche potential funnels the wind material toward the companion star. This can result in accretion efficiencies well beyond traditional Bondi-Hoyle-Lyttleton wind accretion \citep{Hoyle1939,Bondi1944,Mohamed2007}. The maximum efficiency of the WRLOF is likely around 50\% \citep{Abate2013}, where WOCS 4540 requires an efficiency of at least 60\% to explain the luminosity and derived mass (Figure~\ref{fig:stability}). It is unlikely that WRLOF is solely responsible for the mass-transfer history of WOCS 4540.  

We conclude that WOCS 4540 was initially an almost equal-mass binary with a primary mass of 1.2 \msun\ and a secondary mass of 1.1--1.15 \msun. The system experienced stable Roche lobe overflow mass transfer at periastron passage with a mass transfer efficiency of at least 60\%, resulting in a BSS with a mass of 1.55 \msun. It is possible that the stable mass-transfer events were augmented by WRLOF at farther separations in the eccentric orbit, or earlier in the evolution before the giant star radius reached the Roche lobe radius at periastron. 

We leave the detailed modeling of the binary evolution during mass transfer to later efforts.

\subsection{Formation Pathway for WOCS 5379}
\label{sec:5379}
The formation of WOCS 5379 challenges our current understanding of binary mass transfer. Given the presence of a He-core WD in the system, we know the mass transfer occurred while the initial primary was on the RGB, undergoing Case B mass transfer. 

The photometry of WOCS 5379 is best described by a BSS progenitor of 0.7 \msun\ that accretes 0.57 \msun\ to reach a final mass of 1.27 \msun\  (Figure~\ref{fig:bestfittracks}). The final period may be inconsistent with theoretical predictions for stable mass transfer (Figure~\ref{fig:massperiod}) and an estimate of the adiabatic response of the 1.2 \msun\ giant star to mass transfer also shows the system would be unstable (Figures~\ref{fig:stability} and ~\ref{fig:stabilitywithwind}) under a wide range of assumptions about the mass-transfer efficiency, progenitor mass, and wind mass loss prior to onset of Roche lobe overflow. 

One possible explanation for the above would be if WOCS 5379 experienced a single-binary (1+2) or binary-binary (2+2) interaction post-mass transfer that shrunk the orbit.  To calculate the probability of such an interaction having occurred, we use the same assumptions for the host cluster properties of NGC 188 outlined in Section 3.2 of \citet{Leigh2011} (right-hand column).  We multiply these timescales (i.e., equations A8 and A10 in \citet{Leigh2011}) by the total number of binaries in the cluster to obtain the times for a \textit{specific} binary (i.e., WOCS 5379) to undergo a 1+2 or 2+2 interaction.  This gives 1+2 and 2+2 interaction times of, respectively, approximately 35.3 Gyr and 40.3 Gyr.  These timescales can be used to compute the probability of a 1+2 or 2+2 interaction having occurred post-mass transfer by taking the derived time since mass transfer ended of 250 Myr and dividing by each timescale.  This gives a probability of a 1+2 or 2+2 interaction involving WOCS 5379 having occurred post-mass transfer of, respectively, approximately 0.7\% and 0.6\%.  We conclude that the orbital properties of WOCS 5379 have a negligible probability of being affected by a dynamical interaction, meaning the observed orbit most likely reflects the orbit immediately following mass transfer. 

Traditionally, unstable mass transfer should cause the system to go into a common envelope phase with mass transfer efficiencies of less than 10\% and with short final orbital periods on the order of days \citep{Woods2011}. And yet, the best-fit accretion track for WOCS 5379 requires a very high mass-transfer efficiency of at least 65\% to reach both the final BSS mass and the proximity to the zero-age main sequence necessary to explain the photometry of this system. If the system has a history of unstable mass transfer, it is unclear how enough mass could have been accreted by the BSS progenitor to form WOCS 5379. Additionally, with a period of 120 days, the system is much wider than typical post-common envelope systems with periods of only a few days \citep{Ivanova2013}. If our understanding of unstable mass transfer changes to be consistent with the binary period and mass transfer efficiency of WOCS 5379, this system could place helpful constraints on new unstable mass transfer theories.

This conflict may be evidence that the application of the adiabatic approximation used in Section~\ref{sec:stability} is not valid in this case. Another common criteria for determining whether stable mass transfer can occur is based on the critical mass ratio ($q_c= M_{\text{donor}} / M_{\text{accretor}}$) needed for stable mass transfer. A traditional critical mass ratio formula used in many population synthesis codes \citep{Hjellming1987} yields a critical mass ratio of only $q_c= 0.74$ compared to the $q_c = 1.7$ required for our formation model of a 1.2 \msun~red giant accreting onto a 0.7 \msun~main-sequence star. Updated calculations of $q_c$ yield somewhat higher values. For example, \citet{Woods2011} suggest that including the superadiabatic response to mass loss of the surface layers of an approximately 1.0 \msun\ red giant donor with an approximately 0.4~\msun\ core would yield a $q_c$ of 1.0--1.1. \citet{Chen2008} models for BSS formation yield $q_c$ of approximately 1.0 for systems similar to the model we propose for WOCS 5379. While closer, these criteria for stability still yield a $q_c$ too low for our formation model.

\citet{Pavlovskii2015} compare the adiabatic treatment of giant donor stars to 1D stellar models and adopt a different stability condition for whether there is overflow through the L2/L3 Lagrangian points. Importantly for this work, the L2/L3 overflow criteria can allow for stable mass transfer in binaries that fail the criteria set by the adiabatic response of the giant star. A robust comparison of WOCS 5379 to this criteria would require careful modeling of the binary evolution throughout mass transfer, which is not trivial given the non-zero eccentricity. We believe WOCS 5379 may be an important test case for the L2/L3 stability criterion in future studies.

Given the current orbital eccentricity of this system, it is possible that mass transfer occurred in an initially eccentric system as we also suggest for WOCS 4540. In this case, Roche lobe overflow may only occur during periastron passage. Much is not known about mass transfer in eccentric systems, but it could increase the stability of mass transfer in this system and explain the current non-zero orbital eccentricity. WOCS 5379 and WOCS 4540 might both therefore be good test cases for new prescriptions for eccentric mass transfer (e.g., \citealt{Hamers2019}).

It is possible that our model assumptions of no internal mixing and standard abundances could impact our interpretation of matching the accretion tracks to the observed parameters of WOCS 5379. In particular, if the WOCS 5379 helium abundances are enhanced due to deep mixing it could impact the observed color \citep{Sills2001} and therefore our inferred blue straggler mass. If deep mixing has occurred, the progenitor of WOCS 5379 could be more massive than what is presented here. Modeling the internal mixing, surface abundances, and diffusion of the BSS due to the accretion of donor star material is non-trivial and requires modeling the internal dynamics of the star simultaneously with the binary evolution and mass transfer, including angular momentum changes in the system. The binary evolution and mass transfer is further complicated by the non-zero eccentricity of the system. Although this type of modeling is beyond the scope of this paper, WOCS 5379 will provide an interesting and important test case for future detailed binary mass transfer modeling efforts.  



\section{SUMMARY}
\label{sec:summary}
We analyze COS spectroscopy of two WD companions of BSSs WOCS 4540 and WOCS 5379 in the old open cluster NGC 188. We fit the WD spectra with WD atmospheres across four different distance assumptions, finding that the WD masses and cooling ages are roughly consistent across each distance assumption (Table~\ref{tab:fitresults}). Adopting the \textit{Gaia}-based cluster distance of $1847\pm107$ pc we find that WOCS 4540 has a CO-core WD with \teff=$17000^{+140}_{-200}$ K and \logg=$7.80^{+0.06}_{-0.06}$, corresponding to a WD mass of $0.53^{+0.03}_{-0.03}$ \msun\ and a cooling age of $105^{+6}_{-5}$ Myr. WOCS 5379 has a He-core WD with \teff=$15500^{+170}_{-150}$ K and \logg=$7.50^{+0.06}_{-0.05}$, corresponding to a mass of $0.42^{+0.02}_{-0.02}$ \msun\ and a cooling age of $250^{+20}_{-20}$ Myr. Adopting a current cluster age of 6.2 Gyr, we find that 100--300 Myr ago the cluster turnoff mass was approximately 1.2 \msun. We conclude that both systems began with 1.2 \msun\ primary stars, but different secondary stars and initial binary parameters led to very different BSS products. 

Combining the ages of these systems with constraints derived from membership in NGC 188, we explore possible mass-transfer formation histories using a grid of MESA accretion evolutionary tracks compared to the observed physical parameters of WOCS 4540 and WOCS 5379.  This comparison suggests that WOCS 4540 formed from a 1.2 \msun\ and 1.14 \msun\ progenitor binary that underwent mass transfer while the primary star was on the AGB. The final binary period and the accretion tracks are both consistent with stable mass transfer occurring during periastron. It is possible that stable mass transfer was enhanced by wind Roche lobe overflow events, especially outside of periastron passages. 

WOCS 5379 likely formed from a 1.2 \msun\ and 0.7 \msun\ progenitor system. According to theoretical predictions of mass transfer and across a range of reasonable assumptions regarding the adiabatic response of the giant star to mass loss, WOCS 5379 would not have had a history of stable mass transfer. WOCS 5379 requires relatively high mass transfer efficiency of at least 65\% to recreate both the BSS mass and proximity to the zero-age main sequence, but unstable mass transfer is typically understood to lead to a common envelope with very low mass transfer efficiencies of less than 10\% \citep{Woods2011} and a short-period orbit. Yet WOCS 5379 exists with an orbital period of 120 days. This system could be an important test case for the stability criterion used in Case B mass transfer scenarios. 

WOCS 4540 and WOCS 5379 provide interesting opportunities to study mass transfer binary evolution in detail. Both have non-zero eccentricities, which is a challenge for typical binary mass transfer models that assume that systems circularize when mass transfer begins \cite[e.g.,][]{Hurley2002}. These observations can be used to constrain the parameter space for further theoretical efforts to understand mass transfer physics in detail.

\acknowledgments
We thank the anonymous referee for their thoughtful comments and suggestions that improved this paper. We are grateful to D. Koester for sharing his grid of white dwarf atmosphere models. We thank J. J. Hermes for his insightful comments on the appropriate mass ranges for the WD fits. This paper is based on observations made with the NASA/ESA \textit{Hubble Space Telescope}, obtained at the Space Telescope Science Institute, which is operated by the Association of Universities for Research in Astronomy, Inc., under NASA contract NAS 5-26555. These observations are associated with program \#13354 and R.D.M. has been supported by NASA HST-GO-13354.001-A and NSF AST- 1714506. E.M.L. is funded by a National Science Foundation Astronomy and Astrophysics Postdoctoral Fellowship under award No. AST-1801937.

\facilities{HST (COS)}

\software{MESA \citep{mesa}, 
numpy \citep{numpy}, 
astropy \citep{astropy:2013,astropy:2018}, 
emcee \citep{emcee}, 
seaborn \citep{seaborn}, 
matplotlib \citep{matplotlib}, 
scipy \citep{scipy}, 
ipython \citep{ipython}.}

\bibliography{biblio}

\end{document}